\PassOptionsToPackage{prologue,dvipsnames}{xcolor}
\documentclass[sigconf]{acmart}

\makeatletter
\def\@ACM@checkaffil{
    \if@ACM@instpresent\else
    \ClassWarningNoLine{\@classname}{No institution present for an affiliation}%
    \fi
    \if@ACM@citypresent\else
    \ClassWarningNoLine{\@classname}{No city present for an affiliation}%
    \fi
    \if@ACM@countrypresent\else
        \ClassWarningNoLine{\@classname}{No country present for an affiliation}%
    \fi
}
\makeatother

\AtBeginDocument{%
  \providecommand\BibTeX{{%
    \normalfont B\kern-0.5em{\scshape i\kern-0.25em b}\kern-0.8em\TeX}}}

\copyrightyear{2023}
\acmYear{2023}
\setcopyright{licensedothergov}
\acmConference[KDD '23]{Proceedings of the 29th ACM SIGKDD Conference on Knowledge Discovery and Data Mining}{August 6--10, 2023}{Long Beach, CA, USA}
\acmBooktitle{Proceedings of the 29th ACM SIGKDD Conference on Knowledge Discovery and Data Mining (KDD '23), August 6--10, 2023, Long Beach, CA, USA}
\acmPrice{15.00}
\acmDOI{10.1145/3580305.3599502}
\acmISBN{979-8-4007-0103-0/23/08}

\usepackage{CJKutf8}
\usepackage{graphicx}
\usepackage{caption}
\usepackage{subcaption}  

\usepackage{float}
\usepackage[normalem]{ulem}

\usepackage{algorithm}
\usepackage{makecell}

\usepackage{algpseudocode}
\usepackage{hyperref}
\usepackage{multirow}
\usepackage{tabularx}
\algnewcommand\algorithmicinput{\textbf{Input:}}
\algnewcommand\algorithmicoutput{\textbf{Output:}}
\algnewcommand\Input{\item[\algorithmicinput]}%
\algnewcommand\Output{\item[\algorithmicoutput]}
\algnewcommand\algorithmicforeach{\textbf{for each}}
\algdef{S}[FOR]{ForEach}[1]{\algorithmicforeach\ #1\ \algorithmicdo}

\newcolumntype{m}{>{\hsize=.13\hsize}X}
\newcolumntype{s}{>{\hsize=.09\hsize}X}
\newcolumntype{g}{>{\hsize=.08\hsize}X}
\newcolumntype{b}{>{\hsize=.70\hsize}X}

\newcolumntype{?}{!{\vrule width 1pt}}

\newcommand{\papername}{Shilling Black-box Review-based Recommender Systems through Fake Review Generation}

\usepackage[dvipsnames, prologue]{xcolor}

\definecolor{d_orange}{RGB}{197, 90, 17}
\definecolor{my_blue}{RGB}{0, 176, 240}

\begin{document}
\begin{CJK*}{UTF8}{bsmi}
\title{\papername}



\author{Hung-Yun Chiang }
\affiliation{%
  \institution{National Tsing Hua University}
  \country{Taiwan}
  }
\email{harrychiang0@gmail.com}

\author{Yi-Syuan Chen}
\affiliation{%
  \institution{National Yang Ming Chiao Tung University, Taiwan}
  }
\email{yschen.ee09@nycu.edu.tw}

\author{Yun-Zhu Song }
\affiliation{%
  \institution{National Yang Ming Chiao Tung University, Taiwan}
  }
\email{yzsong.ee07@nycu.edu.tw}

\author{Hong-Han Shuai}
\affiliation{%
  \institution{National Yang Ming Chiao Tung University, Taiwan}
  }
\email{hhshuai@nycu.edu.tw}

\author{Jason S. Chang}
\affiliation{%
  \institution{National Tsing Hua University}
  \country{Taiwan}
  }
\email{jason@nlplab.cc}


\begin{abstract}
Review-Based Recommender Systems (RBRS) have attracted increasing research interest due to their ability to alleviate well-known cold-start problems. RBRS utilizes reviews to construct the user and items representations. However, in this paper, we argue that such a reliance on reviews may instead expose systems to the risk of being shilled. To explore this possibility, in this paper, we propose the first generation-based model for shilling attacks against RBRSs. Specifically, we learn a fake review generator through reinforcement learning, which maliciously promotes items by forcing prediction shifts after adding generated reviews to the system. By introducing the auxiliary rewards to increase text fluency and diversity with the aid of pre-trained language models and aspect predictors, the generated reviews can be effective for shilling with high fidelity. Experimental results demonstrate that the proposed framework can successfully attack three different kinds of RBRSs on the Amazon corpus with three domains and Yelp corpus. Furthermore, human studies also show that the generated reviews are fluent and informative. Finally, equipped with Attack Review Generators (ARGs), RBRSs with adversarial training are much more robust to malicious reviews.

\end{abstract}




\maketitle

\section{Introduction}

\begin{figure}[h]
    \centering
    \includegraphics[width=0.85\columnwidth]{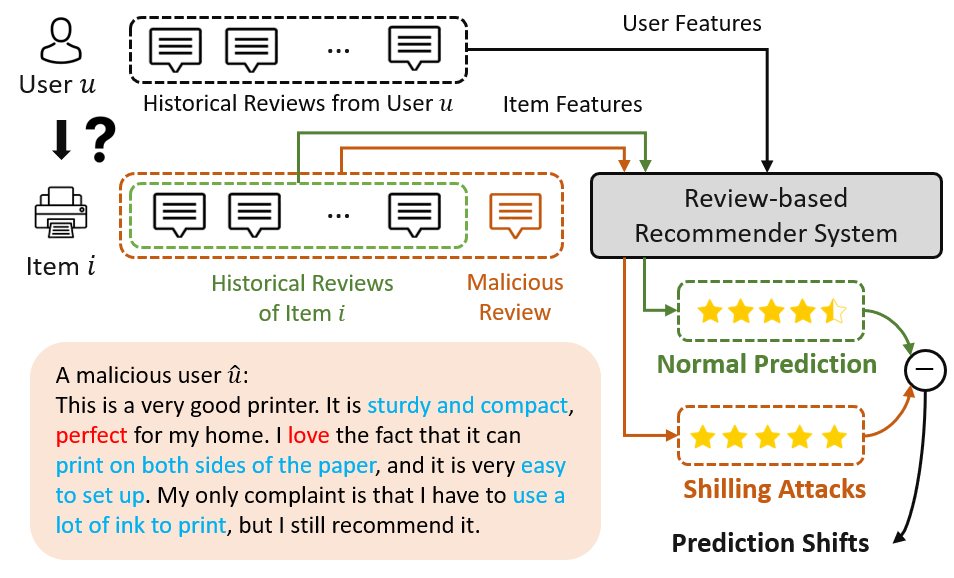} 
    \caption{An illustrative example of a shilling attack by the proposed attack review generator. A malicious review is generated to promote a certain item, i.e., increasing the predicted rating score of an RBRS. The blue words describe the detailed features of the product; while the red words describe the attack keywords against the RBRS.}
    \label{fig1}
    \vspace{-3mm}
\end{figure}

Recommender systems, which predict user preferences from historical records, play a crucial role in helping customers with their decision-making process~\cite{recsys-survey1, recsys-survey2, recsys-survey3}. Recently, a variety of Review-Based Recommender Systems (RBRS) have been proposed to incorporate textual reviews from users for user modeling and recommendation processes~\cite{deepconn, AHN, RMG, NRCMA, RGCL, narre, nrpa}, and have been applied to real-world e-commerce systems~\cite{case-amazon}. For instance, Zheng et al.~\cite{deepconn} is the first work that uses convolutional neural networks (CNNs) to encode user and item review information, while subsequent works further leverage advanced techniques such as attention mechanisms~\cite{transformers} to encode review information~\cite{NRCMA, AHN, narre} or construct a graph structure for modeling user-item interactions~\cite{RGCL, RMG}.

Although introducing reviews into recommendation models has been proven useful for alleviating well-known cold-start problems and sparse interactions, this paper argues that textual reviews also make RBRSs vulnerable to \textit{malicious attacks}. In other words, the recommendation results can be manipulated deliberately by adding malicious item reviews, which is also known as a \textit{Shilling Attack}~\cite{shilling-nips, shilling-in-practice, shilling-4}. Specifically, the shilling attack aims to promote certain items (so that items are recommended more frequently) or de-promote them (so that items are recommended less often) by adding the poisoned reviews. Figure~\ref{fig1} illustrates an example of a shilling attack. RBRS uses user and item reviews to predict the rating between the unseen user-item pair. In a shilling attack, an additional fake review (hereafter referred to as the attack review) is added to the item's historical reviews to promote item $i$. After the RBRS takes the attack review, the rating is increased from 4.5 to 5 stars, which is considered a successful shilling attack to promote the item for this user. With the observations of the vulnerability to attack reviews, we focus on the impacts of attack reviews on the RBRSs and how to defend against the attack reviews. 

It is worth noting that a related line of research to defend against attack reviews is to utilize users' historical interaction records for detecting malicious users and further removing their reviews~\cite{detect-user, detect-user1}. However, it is easy to create new account or buy existing normal accounts to add malicious reviews. In this paper, we focus on generating the attack reviews automatically and improve the model robustness. To the best of our knowledge, this is the first work to study how to attack RBRSs by generating \textit{new} reviews. An intuitive way to perform the shilling attack on RBRSs is to create a new user and post new reviews by copying the one-star (de-promote) or five-star (promote) reviews. However, copying other users’ reviews may i) violate the regulations of review platforms and ii) be easily detected and removed~\cite{fake-review-detect-1, fake-review-detect-2}. As such, to successfully perform a shilling attack on RBRSs, we propose leveraging a learning approach for generating new shilling reviews instead of directly copying existing reviews.

Nevertheless, generating reviews for shilling attacks is challenging due to the following three challenges. 1) The attack reviews are generated without ground truth. Different from review generation~\cite{selsum, Fewsum, ni-mcauley-2018-personalized}, there is no ground truth for supervising the generation process to attack RBRSs. One possible approach is to leverage existing textual adversarial attacks~\cite{textbugger, unitrigger, textattack} proposed for other applications. However, this kind of textual adversarial attack is performed by changing, inserting, and deleting words or tokens in existing sentences. As such, the generated reviews may still be similar or even identical to the existing reviews. 2) Attack reviews should not only look realistic (fidelity), but also be able to attack RBRS (effectiveness). This is challenging because there is a tradeoff between making an attack review effective and making it more realistic. For example, a review of ``Fantastic, fantastic, fantastic, fantastic movie’’ may be effective due to the repetitive keywords for RBRSs but is less realistic. In contrast, a realistic usually contains vivid descriptions of the items while they are ineffective for promoting/de-promoting them. 3) The generated reviews should be different for items in the same category (diversity). Since the weakness of RBRSs may be some keywords, it is easy to generate the same reviews for different items, e.g., \textit{``This product is very good, it's great, I recommend it''}. However, to mimic normal users who usually leave different reviews and provide abundant and detailed information, it is desirable to generate reviews with high variety and detailed item information.

To address these challenges, in this paper, we propose an Attack Review Generator (ARG) to perform shilling attacks against RBRSs. ARG adopts a generative model to produce completely new reviews based on current existing reviews, which is equivalent to creating a new user bot to leave a new item review in a real recommendation scenario. To address the first issue, we introduce reinforcement learning~\cite{rl-williams, seq2seq_RL_1, seq2seq_RL_2} and use a prediction shift~\cite{ps-paper-1, ps-paper-2, ps-paper-3} as one of our rewards. The prediction shift summarizes the differences between the predicted rating of an item \textit{among all users} before and after adding the attack review, which measures the effectiveness of the generated attack reviews on the RBRS. Moreover, to address the second challenge, we introduce a pre-trained language model~\cite{gpt2} to measure the fluency of the generated reviews as another reward for RL and make RL learn to strike a good balance between fidelity and effectiveness. Finally, to address the third challenge, as previous works~\cite{RGCL, uarm, aspect_paper_1} have shown that human-written reviews always contain fine-grained aspect information\footnote{For example, one review says ``\textit{This product has good quality}'', and another says ``\textit{This product is so sturdy, nice quality}''. They are both describing the item as being good quality, but the second review mentions a concrete property which is the fine-grained information from the aspect of quality.}, we use ROUGE scores~\cite{rouge} of product attributes as another RL reward to encourage ARGs to capture product attributes in the existing reviews. Moreover, we additionally propose an aspect generation loss to encourage ARG to generate more fine-grained information for the corresponding attributes.

The main contributions are summarized as follows.
\begin{itemize}
    \item To the best of our knowledge, this is the first work that uses a generative model to perform a shilling attack on RBRSs. We further demonstrate that RBRS can be more robust to malicious attacks with adversarial training.
    \item We propose the Attack Review Generator (ARG), which can effectively generate highly relevant and high-quality reviews for specific items and attacks.
    \item Experiments on public datasets show that ARG successfully attacks three different types of RBRSs and outperforms all the baselines by at least 40.5\% and 82.6\% in terms of prediction shift on Amazon and Yelp, respectively.
\end{itemize}

\section{RELATED WORK}
\subsection{Attack on a Recommender System}
Previous studies~\cite{shilling-nips, shilling-in-practice, shilling-3-user-profile,shilling-5-user-profile, black-box-1, shilling-incomplete, shilling_user_purchase_2, kdd-attack-against-recsys-2022, kdd-rebuttal-shilling} have shown that the shilling attack, which injects malicious data into the RS, is effective. For instance, several works propose to inject user purchase records~\cite{KDD-triple-fake-user-poisoning, KDD-knowledge-bbox-fake-user, shilling_user_purchase_1, shilling_user_purchase_2, shilling_user_purchase_3, shilling_user_purchase_4, shilling_user_purchase_5, shilling_user_purchase_6} or user profiles~\cite{shilling-3-user-profile, shilling-5-user-profile} to disrupt the RS output and control which items are pushed. Lin et al.~\cite{shilling-3-user-profile} and Chen et al.~\cite{shilling-5-user-profile} both generated fake user profiles with adversarial training to poison the RS. Cohen et al.~\cite{black-box-1} attacked an image-based recommender system that uses product images in the recommendation process and performs image adversarial attacks~\cite{image_adv_1, image_adv_2}, to achieve shilling RSs. Zhang et al.~\cite{shilling-incomplete} identified shilling attacks with incomplete or even perturbed user-item interaction attacks. To defend RSs, several works have been proposed to detect fake data~\cite{shilling-nips, fake-review-detect-1, fake-review-detect-2, fake-review-detect-3}. For instance, Pang et al. proposed unorganized malicious attacks detection algorithm, where attackers separately utilize a few user profiles to attack different items without organizer. Compared with previous works, this is the first work to study how to attack RBRSs by automatically generating \textit{new} reviews.

\subsection{Review-based Recommendation System}
Review-based Recommender Systems aim to learn the representation of users and items from their historical review texts for estimating the user ratings of items. For example, DeepCoNN~\cite{deepconn} learns user and item embeddings by concatenating historical reviews as review documents and feeding them into the text-CNN network. Moreover, many following studies~\cite{narre, daml, AHN, NRCMA} use different attention mechanisms to capture more information between users and items' historical reviews. For example, Chen et al.~\cite{narre} proposed a review attention mechanism where each review is given a different attention weight according to the importance. Liu et al.~\cite{daml} proposed dual attention mutual learning, which contains a local attention layer for selecting informative words from reviews and a mutual attention layer for learning the relevant semantic information between user reviews and item reviews. On the other hand, another line of studies has attempted to introduce graph structures to recommendation systems. For example, RMG~\cite{RMG} models reviews and rating interactions through the graph structure with a multiview framework. Furthermore, SSG~\cite{ssg} uses the multiview approach with sequences and graphs to capture long-term and short-term features from users and items. RGCL~\cite{RGCL} further introduces contrastive learning of graphs in user-item interaction graphs that solves the limited user interaction behaviors. Most RBRSs utilize reviews during the inference stage to capture the latest information from users and items. However, this behavior of RBRS also exposes them to the risk of being attacked. 

\section{Problem Formulation}
\textbf{Review-based Recommendation System.} 
In a review-based recommendation scenario, there are $N$ users and $M$ items. We denote the interactions between users and items by a rating matrix $R \in \mathbb{R}^{N \times M}$, where $R_{u i} \in\{1,2,3,4,5\}$ indicates the rating of item $i$ given by user $u$. Moreover, $S_{ui}$ represents a review from user $u$ to item $i$. The goal of an RBRS is to predict the rating score between unseen user-item pairs based on their historical reviews as follows:
\begin{equation}
 {R}_{u i} = RBRS(E_u, E_i, S_{u}, S_{i}),
\end{equation}
where $E_u$ and $E_{i}$ are the ID embeddings of the corresponding user and item, respectively. Furthermore, $S_u$ represents the set of historical reviews given by user $u$ and $S_i$ represents the set of historical reviews received by item $i$.

\noindent\textbf{Attack Scenario.} 
In this work, we aim to explore the model capabilities of performing a shilling attack on Review-Based Recommendation Systems (RBRSs). The shilling attack uses an attack review $S_{\hat{u}i}$ written by a malicious user bot $\hat{u}$ to change the RBRS prediction of a specific item $i$. For an attack review, the performance is evaluated by the \textit{prediction shift}~\cite{ps-paper-1, ps-paper-2, ps-paper-3} of RBRS, which is the rating difference before and after adding the review attack among all candidate users. Specifically, let $\hat{R}_{u i}$ and $PS_{u}(S_{\hat{u}i})$ respectively denote the new rating score after combining the attack review $S_{\hat{u} i}$ with $S_i$ and the prediction shift caused by $S_{\hat{u}i}$ on a single user $u$.
The \textit{prediction shift} can be expressed as follows:
\begin{equation}
\label{ps}
  PS_{u}(S_{\hat{u}i})  = \hat{R}_{u i} - {R}_{u i},
\end{equation}

\begin{equation}
\label{attack-eq}
  \hat{R}_{u i} = RBRS(E_u, E_i, S_{u}, S_{i} \cup S_{\hat{u}i}).
\end{equation}
Our goal is to maximize $PS(S_{\hat{u}i})$, which represents the average prediction shift among all candidate users while keeping the attack review unnoticeable. In addition, our attack is under a black-box setting to make the attack scenario more realistic and in line with the real world. In other words, only the output predictions of RBRS are accessible during our model training, while the parameters and configurations are unknown.

\begin{figure*}[!t]
    \centering
    \includegraphics[width=0.7\textwidth]{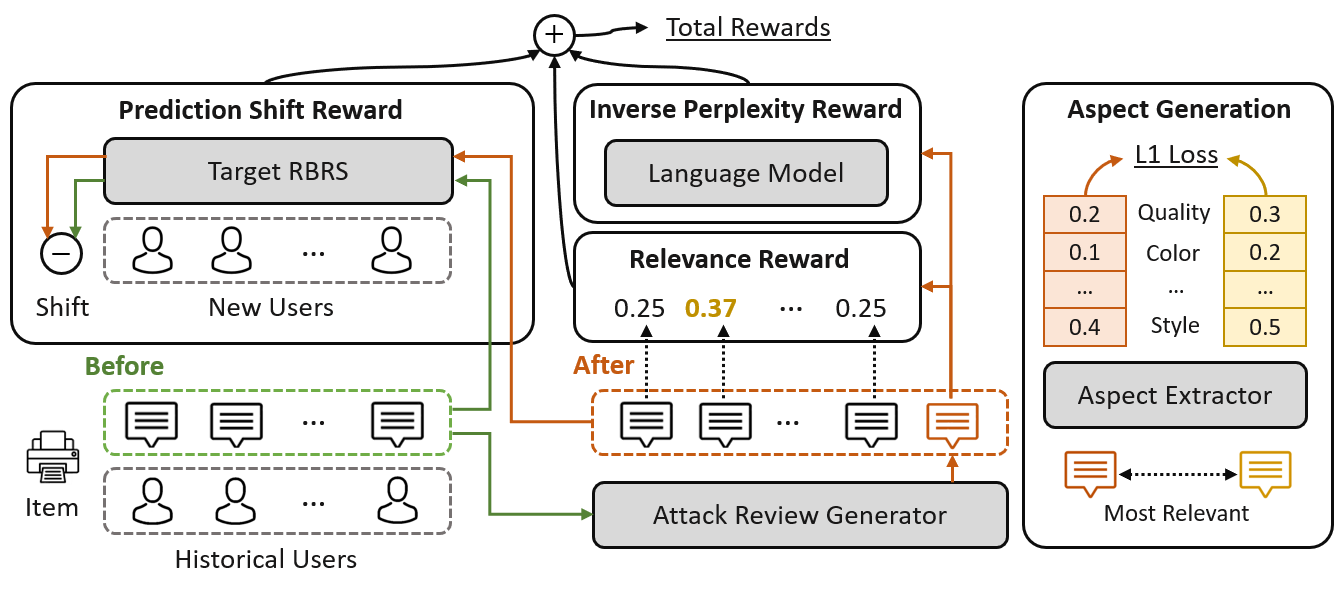}
    \caption{The training framework of ARG.}
    \label{fig:framework}
\end{figure*}

\section{METHODOLOGY}

In this section, we present the details of each module in the proposed methods. Sec.~\ref{sec:arg} describes the architecture and pre-training scheme of the Attack Review Generator (ARG). Sec.~\ref{sec:rl} elaborates on how to apply reinforcement learning to train the ARG while there is no supervision for the attack reviews. Sec.~\ref{aspect-section} further introduces an aspect generator to enhance the aspect information of generated reviews based on the observation that human reviews tend to include more aspect information. Sec.~\ref{sec:pipeline} presents the overall training pipeline of the ARG.


\subsection{Attack Review Generator}
\label{sec:arg}
Given an item, the goal of the Attack Review Generator (ARG) is to generate a corresponding review such that the recommendation results from the RBRSs can be shifted. We use transformer-based encoder-decoder architecture~\cite{transformers} as the base model with pre-trained weights. To generate high-quality reviews and alleviate the cold-start reinforcement learning problem~\cite{cold-start-rl}, the ARG is first pre-trained to generate a regular review as a warm-up. Specifically, we create a self-supervised dataset and train the model in a leave-one-out~\cite{Fewsum} manner:

\begin{equation} \label{leave-one-out}
    \begin{aligned} \theta^{*} =\underset{\theta}{\arg \max } \frac{1}{J I} \sum_{j=1}^J \sum_{i=1}^I \log G_\theta\left(r_i^j \mid E_\theta\left(r_{-i}^j\right)\right), \end{aligned}
\end{equation}
where $E_\theta$ and $G_\theta$ represent the encoder and decoder parameters respectively, while $J$ and $I$ denote the number of groups and the number of reviews in a group, respectively, and the reviews in the same group describe the same item. During the pre-training stage, we train a conditional language model by taking a randomly selected review $r_{i}$ as a pseudo label. Specifically, the model is learned to generate $r_{i}$ based on the remaining reviews $r_{-i}$. With leave-one-out pre-training, our model can generate a new review based on the historical reviews of any item and can capture the common information of a specific item.

\subsection{Reinforcement Learning for Attack}
\label{sec:rl}
After leave-one-out pre-training, the ARG is only able to generate a fake review for a certain item, but the fake review may not be a threat to the RBRS. To generate reviews that can effectively attack the RBRS, we need to further train the ARG to make prediction after the pre-training stage. However, we encounter the following two problems: 1) there is no ground truth to train ARGs for generating reviews that can produce a larger prediction shift, and 2) the prediction shift is a non-differentiable metric~\cite{regen}. Therefore, we leverage reinforcement learning \cite{rl-williams} framework with custom rewards to overcome both of these problems. The proposed ARG is considered an agent while the action is token selection at each generation step. The ARG $\theta$ is learned by maximizing the expected value of rewards, and the loss function can be formulated as:

\begin{equation}
    \begin{aligned}
{L}_{\mathrm{RL}} &=-\sum_{s_0, \ldots, s_\ell} p_\theta\left(s_0, \ldots, s_\ell\right) R\left(s_0, \ldots, s_\ell\right) \\
&=-\mathbb{E}_{\left[s_0, \ldots, s_\ell\right] \sim p_\theta} R\left(s_0, \ldots, s_\ell\right) \\
&=-\mathbb{E}_{S \sim p_\theta} R\left(S\right),
\end{aligned}
\end{equation}
where $S = S_{\hat{u}i} = (s_0, s_1, ..., s_\ell)$ is the generated review sequence and $R(S)$ is the reward of $S$. We obtain the minimize loss function $L_{RL}$ by taking the negative value of the exception. Then, we use the policy gradient \cite{policy-gradient} method to update the ARG:
\begin{equation}
    \nabla_\theta L_{\mathrm{RL}} \propto-\left(R\left(S\right)-b\right) \nabla_\theta \log p_\theta\left(S\right),
\end{equation}
where $b$ denotes a baseline for reducing gradient estimator variance. Following Self-Critical Sequence Training (SCST)~\cite{scst, petgen}, we take the greedy output $R(S^\ast)$ as the reward baseline. That is, for each generation step, $S^\ast$ takes the token with the highest probability the output. The loss of SCST can be calculated as follows:

\begin{equation}
\label{loss-scst}
    \nabla_\theta L_{\mathrm{SCST}} \propto-\left(R\left(S\right)-R\left(S^*\right)\right) \nabla_\theta \log p_\theta\left(S\right),
\end{equation}
where $S$ is the sampled sequence, and $S^*$ is generated by greedy search from the ARG. The proposed three rewards in self-critical sequence training will be introduced in the following subsections.

\subsubsection{Prediction Shift Reward}
\label{reward-ps}
The effectiveness of a generated review $S_{\hat{u}i}$ is considered to be the prediction difference before and after an RBRS considering the review. To generate the reviews that can attack the RBRS, we directly take the prediction shift of the RBRS as one of the rewards. Specifically, the reward is obtained from $PS(S_{\hat{u}i})$ as shown in Equation~\ref{ps}.

\subsubsection{Inverse Perplexity Reward}
\label{reward-invppl}
To make the generated reviews look realistic, we introduce \textit{perplexity} as one of our rewards. Perplexity is a metric for evaluating Language Models (LMs)~\cite{gpt2}. We use a well-trained LM that is fine-tuned on our review corpus to evaluate the perplexity of attack reviews: 

\begin{equation}
    PPL(S)=\exp \left( -\frac{1}{\ell} \sum_{i=1}^\ell \log p_\theta\left(s_i \mid s_{<i}\right) \right),
\end{equation}
where $S = S_{\hat{u}i} = (s_0, s_1, ..., s_\ell$) and $s_l$ is a generated token. A lower perplexity implies a higher log-likelihood of $S_{\hat{u}i}$, which means that $S_{\hat{u}i}$ is more similar to the reviews in our corpus. Therefore, we take the inverse perplexity as the reward:
\begin{equation} \label{inv-ppl}
    InvPPL(S) = \frac{1}{PPL(S)}.
\end{equation}

\subsubsection{Relevance Reward}
\label{reward-relevance}

To avoid irrelevant content, the model should generate reviews consistent with the rest of the reviews or related to the item property. We propose applying the ROUGE~\cite{rouge} score as a reward to encourage ARG to capture more information from $S_{i}$:
\begin{equation}
\label{relevance}
    Relevance(S) = \frac{1}{H} \sum_{h=1}^H ROUGE \left(S, S_{{u_{h}}i}\right),
\end{equation}
where $S_{{u_{h}}i}$ denotes the $h$-th review in $S_i$, written by $u_h$ which has interaction records with $i$. This reward encourages our model to capture the common information in $S_i$, which is the item information and its features. The final rewards are the summation of the above three rewards\footnote{In our experiments, different weightings do not strike a better balance. As the scale of three rewards is similar, we set the weightings equal and leave the weighting mechanisms for future work.}:
\begin{equation}
    \label{rewards-final}
    R(S) = PS(S) + InvPPL(S) + Relevance(S).
\end{equation}

\subsection{Aspect Generation}
\label{aspect-section}

Previous works~\cite{RGCL,uarm,aspect_paper_1} have shown that human-written reviews often contain fine-grained aspect information, providing more details about product attributes. To make the generated attack reviews more realistic and diverse, it is desirable if ARG can generate aspect information in the attack reviews. However, aspect information may not be included in every dataset for training. Thus, we apply a state-of-the-art aspect extractor ABAE~\cite{abae} as our aspect extraction model, which can learn the hidden aspects in the training corpus in an unsupervised way. Accordingly, if there are $K$ different aspects, the aspect distribution $\phi^S \in \mathbb{R}^{K \times 1}$ of an input review can be formulated as:
\begin{equation}
    \phi^S=A B A E(S),
\end{equation}
where the value of each dimension in $\phi^S$ indicates the importance of the corresponding hidden aspect. To construct a learning target for the aspect distribution of generated attack review $S_{\hat{u}i}$, we use the review that most matches the current attack review from the item's historical reviews $S_i$ and build the learning target $\phi^{S_i}_{target}$ as follows:
\begin{equation}
\begin{gathered}
    \phi^{S_{i}}_{target} = ABAE(S_{u_{h}i}^\ast),\\
    S_{u_{h}i}^\ast = \mathop{\arg\max}\limits_{h \in H} ROUGE(S_{\hat{u}i}, S_{u_{h}i}),
\end{gathered}
\end{equation}
where $S_{u_{h}i}^\ast$ represents a selected review from $S_{i}$ that has the highest ROUGE score with the current attack review $S_{\hat{u}i}$ and $H$ denotes the set of users who have written reviews on item $i$. Using reviews with high similarity as a learning target can stabilize the training process since boosting the aspect information on reviews with shared information is easier than that on non-related reviews. Finally, the aspect generation loss is an L1 loss as follows:
\begin{equation}
\label{aspect-loss}
    L_{aspect} = \sum_{k=1}^{K}|\phi^{S_{i}}_{target_k} - \phi^{S_{\hat{u}i}}_k|.
\end{equation}

\subsection{Training Pipeline}
\label{sec:pipeline}

\begin{algorithm}[t]

\caption{Training Pipeline of ARG}

    \begin{algorithmic}[1]
        \Input Historical reviews $S_i$ of item $i$ and the attack target RBRS;
        \Output An attack review $S_{\hat{u}i}$ describing the item $i$;
        \State Fine-tune pre-trained LM and train an ABAE on the reviews corpus;
        \State Pre-train the ARG with leave-one-out objective (Eq.~\ref{leave-one-out});
        \ForEach {$t$ \textbf{in} training steps}
        \State Generate $S_{\hat{u}i}^t$ based on $S_i$ by ARG
        \State Get $PS_t$ for $S_{\hat{u}i}^t$ (Eq. \ref{ps}, \ref{attack-eq})
        \State Get $Inv PPL_t$ for  $S_{\hat{u}i}^t$ (Eqs. \ref{inv-ppl})
        \State Get $Relevance_t$ for $S_{\hat{u}i}^t$ (Eq. \ref{relevance})
        \State $R(S_{\hat{u}i}^t) \gets PS(S_{\hat{u}i}^t) + InvPPL(S_{\hat{u}i}^t) + Relevance(S_{\hat{u}i}^t)$
        \State $R(S_{\hat{u}i}^{t^*}) \gets PS(S_{\hat{u}i}^{t^*}) + InvPPL(S_{\hat{u}i}^{t^*}) + Relevance(S_{\hat{u}i}^{t^*})$

        \State Calculate $L_{SCST}$ with $R(S_{\hat{u}i}^t)$ and $R(S_{\hat{u}i}^{t^*})$ (Eq. \ref{loss-scst})
        \State Calculate $L_{aspect}$ (Eq. \ref{aspect-loss})
        \State Update parameters of ARG with the loss in Eq. \ref{loss-total}
        \EndFor
    \end{algorithmic}
    \label{alg:alg1}
\end{algorithm}

Algorithm~\ref{alg:alg1} details the full ARG training pipeline. First, we have the target attack model: a trained RBRS (black-box), as well as the aspect extractor ABAE and the fine-tuned LM trained in our review corpus. Then, at each training step, the ARG generates the attack review, we use the rewards mentioned in ~\ref{reward-invppl}, \ref{reward-ps} and \ref{reward-relevance} to obtain $L_{SCST}$, and then we obtain $L_{aspect}$ from \ref{aspect-section} to incorporate fine-grained aspects into the reviews. The final optimization goal $L_{total}$ for each step is the weighted sum of these two losses, i.e.,

\begin{equation}
    L_{total} = \lambda L_{SCST} +  (1 - \lambda) L_{aspect},
    \label{loss-total}
\end{equation}
where $\lambda$ is a weighting hyperparameter.

\section{EXPERIMENTS}

\begin{table*}[t]
\caption{Attack performance of different RBRSs trained on different datasets. Without Attack means we evaluated the metrics on the testing set of reviews corpus used as a reference, indicating the performance of the metrics on real reviews. ``Rel'' and  ``\% of AW'' are two metrics that refer to relevance and the percentage of aspect words, respectively.}
    \scriptsize
    \centering
    \begin{tabular}{cc|cccc|cccc|cccc|cccc}
    \toprule
    \multirow{2}{*}{RBRS} & \multirow{2}{*}{Method} & \multicolumn{4}{c|}{\textbf{Musical Instruments}} & \multicolumn{4}{c|}{\textbf{Video Games}}  & \multicolumn{4}{c|}{\textbf{Office Products}} & \multicolumn{4}{c}{\textbf{Yelp}} \\

     & & PS & PPL & Rel & \shortstack{\% of AW}  & PS & PPL & Rel & \shortstack{\% of AW} & PS & PPL & Rel & \shortstack{\% of AW} & PS & PPL & Rel & \shortstack{\% of AW}\\
    \midrule
    
    \multicolumn{2}{c|}{Without Attack (Human Written)} & - & 6.89 & 0.179 & 10.81 & - & 6.64 & 0.175 & 9.57 &  - & 5.67 & 0.172 & 16.44 &  - & 10.45 & 0.228 & 10.09 \\

    \hline
    \multirow{5}{*}{DeepCoNN} 
                          & Copycat & 0.012 & \textbf{6.72} & 0.201 & 10.19 & 0.015 & 6.75 & 0.170 & 10.31 &  0.012 & \textbf{5.84} & 0.204 & 15.99 & 0.029 & \textbf{7.72} & 0.232 & 9.80\\
                          & TextBugger & 0.017 & 36.75 & 0.160 & 9.97 & 0.017 & 37.03 & 0.160 & 9.41 & 0.019 & 38.52 & 0.190 & 13.33 & 0.022 & 37.71 & 0.207 & 8.87\\
                          & HotFlip & 0.035 & 7.22 & 0.211 & 10.04 & 0.019 & 7.12 & 0.167 & 9.89 &  0.026 & 6.03 & 0.201 & 14.47 & 0.030 & 11.33 & 0.219 &  9.93\\
                          & ARG w/o aspect & \textbf{0.391} & 8.32 & 0.257 & 8.21 & 0.401 & 8.09 & 0.249 & 8.53 & 0.194 & 5.95 & 0.228 & 14.12 & \textbf{0.233} & 9.64 & 0.239 & 10.02\\
                          & ARG & 0.363 & 6.96 & \textbf{0.262} & \textbf{14.34} & \textbf{0.431} & \textbf{5.40} & \textbf{0.251} & \textbf{12.29} &  \textbf{0.290} & 6.80 & \textbf{0.235} & \textbf{21.09} & 0.228 & 9.46 & \textbf{0.241} & \textbf{13.12}\\
                          
    \hline
    \multirow{5}{*}{NRCMA} 
                          & Copycat & 0.027 & 7.01 & 0.201 & 10.31 & 0.021 & 6.68 & 0.174 & 9.67 &  0.027 & 5.76 & 0.195 & 15.43 & 0.033 & 9.48 & 0.219 & 9.93\\
                          & TextBugger & 0.043 & 64.62 & 0.187 & 9.89 & 0.032 & 28.16 & 0.161 & 8.17 &  0.033 & 42.06 & 0.187 & 16.72 & 0.041 & 41.39 & 0.197 & 8.54 \\
                          & HotFlip & 0.055 & 7.48 & 0.209 & 9.92 & 0.041 & 7.03 & 0.189 & 9.44 &  0.037 & 6.29 & 0.194 & 16.33 & 0.052 & 10.42 & 0.210 & 8.73\\
                          & ARG w/o aspect & 0.077 & \textbf{5.89} & 0.268 & 9.73 & \textbf{0.091} & 4.85 & \textbf{0.262} & 7.31 &  0.046 & \textbf{4.58} & \textbf{0.267} & 15.31 & \textbf{0.095} & \textbf{8.77} & 0.247 & 10.12\\
                          & ARG & \textbf{0.084} & 6.25 & \textbf{0.277} & \textbf{14.17} & 0.084 & \textbf{4.10} & 0.255 & \textbf{10.47} &  \textbf{0.052} & 4.71 & 0.261 & \textbf{18.74}  & 0.087 & 9.06 & \textbf{0.256} & \textbf{14.49}\\
                          
    \hline   
    \multirow{5}{*}{RMG} 
                          & Copycat & 0.014 & 6.84 & 0.201 & 11.21 & 0.023 & 6.98 & 0.171 & 9.89 &  0.008 & 5.86 & 0.208 & 15.19 & 0.022 & 10.13 & 0.230 & 9.97\\
                          & TextBugger & 0.031 & 61.42 & 0.187 & 10.05 & 0.026 & 24.37 & 0.161 & 8.72 &  0.017 & 39.16 & 0.182 & 14.77  & 0.023 & 55.13 & 0.199 & 9.41\\
                          & HotFlip & 0.055 & 7.37 & 0.207 & 11.81 & 0.078 & 7.41 & 0.179 & 10.12 &  0.020 & 6.16 & 0.223 & 14.97  & 0.026 & 12.08 & 0.223 & 10.03\\
                          & ARG w/o aspect & 0.089 & 7.40 & 0.263 & 10.39 & 0.187 & \textbf{5.07} & \textbf{0.262} & 8.31 &  \textbf{0.040} & 5.19 & \textbf{0.264} & 15.10  & 0.063 & 10.15 & 0.261 & 10.45\\
                          & ARG & \textbf{0.091} & \textbf{6.32} & \textbf{0.267} & \textbf{14.15} & \textbf{0.192} & 5.32 & 0.257 & \textbf{10.48} &  0.037 & \textbf{4.01} & 0.263 & \textbf{17.89} & \textbf{0.079} & \textbf{10.09} & \textbf{0.262} & \textbf{15.96}\\
    \bottomrule  
    \end{tabular}
    \label{tab:main}
    \vspace{-4mm}
\end{table*}

In this section, we conduct extensive experiments to answer the following research questions.
\begin{itemize}
    \item RQ1: Can our work successfully attack the review-based recommender systems under a black-box setting?
    \item RQ2: In addition to attack performance, what is the quality of the generated reviews, especially for fluency and relevance to item information?
    \item RQ3: Is the review generated by ARG realistic enough from a human perspective and AI-written text detectors?
    \item RQ4: Does the RBRS become more robust after adversarial training?
\end{itemize}

\subsection{Experiment Setup}

\noindent\textbf{Datasets}. We evaluate our model on four datasets. Three of them come from three different domains on Amazon 5-score datasets\footnote{https://nijianmo.github.io/amazon/index.html}: \textit{Musical Instruments}, \textit{Video Games}, \textit{Office Products}, while another dataset is from \textit{Yelp}\footnote{\url{https://www.yelp.com/dataset}}. Following previous RBRS works~\cite{daml,RGCL}, we use 10-fold cross-validation, where each dataset is randomly split into training/validation/testing sets with ratios of 80\%/10\%/10\%, respectively. Note that we use the same training, validation and testing datasets for baselines and our model.

\noindent\textbf{Target Review-based Recommender System}. 
To understand the generalizability of our method to various RBRSs, we select three different kinds of RBRSs as the attack targets, including convolutional neural networks (DeepCoNN~\cite{deepconn}), attention mechanisms (NRCMA~\cite{NRCMA}), and graph neural networks (RMG \cite{RMG}), as follows.
\begin{itemize}
    \item \textbf{DeepCoNN} \cite{deepconn} is a well-known baseline in RBRS that uses review documents from users and items as features and applies Text-CNN-based deep neural networks for prediction.
    
    \item \textbf{NRCMA} \cite{NRCMA} is the state-of-the-art RBRS that introduces a two-tower style cross-modality mutual attention mechanism that bridges the user encoder and item encoder to promote information exchange between reviews and ratings. 
    
    \item \textbf{RMG} \cite{RMG} is the first RBRS that proposes leveraging user-item graphs with graph neural networks to better capture user-item interactions for recommendation.
\end{itemize}

It is worth noting that we do not perform attacks on RBRSs that rely solely on user and item embeddings, e.g., \cite{RGCL}, since this kind of approach may be easily out-of-date without retraining. For instance, if an item is found to have a significant defect or its company is reported to have some bad news after RBRS training, the model can not consider the new corresponding reviews but only use the out-of-date user and item embedding for prediction. 

\noindent\textbf{Baselines.} 
Since none of the previous work studies review attacks, we have selected the following three representative works for textual adversarial attacks as baselines. 

\begin{itemize}
    \item \textbf{Copycat (High Rating Review)}: Following the same setting as~\cite{petgen}, for each item $i$, we randomly select a 5-star review in $S_i$ as $S_{\hat{u}i}$. This baseline evaluates the performance that a positive review written by real people can achieve. 
    
    \item \textbf{TextBugger} \cite{textbugger}: TextBugger is a text adversarial attack framework that can attack both white-box and black-box settings. TextBugger uses a modification-based approach to replace/insert/remove tokens in the reviews to deviate the model prediction from the ground truth. The input of the TextBugger is the same as Copycat.
    \item \textbf{HotFlip} \cite{hotflip}: HotFlip originally detects the most significant word in a review based on the gradient of each input token and replaces it with a similar word to increase the loss for performing adversarial attack. In our attack scenario, the prediction shift is used as the loss function to calculate the gradient and replace the word. Similarly, the input for the HotFlip is the same as Copycat.
\end{itemize}

\noindent\textbf{Evaluation Metrics.} 
The following four metrics are calculated to measure the attack review quality.

\begin{itemize}
    \item \textbf{Attack Effectiveness}: Following previous studies~\cite{ps-paper-1, ps-paper-2, ps-paper-3}, we use the prediction shift to measure the efficacy of the attack methods:
    \begin{equation}
    \label{ps-eval-equation}
        PS = \frac{1}{Q} \sum_{i \in I_Q} \frac{1}{ \left\| U_{i} \right\|} \sum_{u \in U_{i}} \hat{R}_{u i} - {R}_{u i},
    \end{equation}
    where $U_i$ denotes the candidate user set with the users who have not rated item $i$. For each attack framework, we use their method to attack the RBRS by selecting $Q$ items as a subset (denoted by $I_Q$) in the testing set to generate attack reviews, and then obtain the average prediction shift for each item. Here, $Q$ is set to 500 for the evaluation. 
    
    \item \textbf{Generation Fluency}: We introduce a well-trained Language Model (LM)~\cite{gpt2} to measure perplexity as a text fluency metric. A lower perplexity score means that the input text is more likely to exist in the review corpus, that is, more likely to be a real review.
    
    \item \textbf{Relevance}: To measure whether the generated review $S_{\hat{u}i}$ captures enough information from item $i$, we use the ROUGE~\cite{rouge} score as a relevance metric:
    \begin{equation}
        \label{rouge-eval}
        \text{Product Relevance} = \frac{1}{\left\| S{_i} \right\|} \sum_{S \in S_{i}} \text{ROUGE-1}(S_{\hat{u}i}, S).
    \end{equation}
    A higher relevance score $S_{\hat{u}i}$ indicates that more information is contained in the generated review.
    \item \textbf{Aspect Performance}: To evaluate whether the generated reviews $S_{\hat{u}i}$ contain the aspect information, we use the same metric used in \cite{uarm}-\textit{Percentage of Aspect Words}. For each aspect generated by ABAE, we calculate the top-30 nearest words for each aspect.\footnote{These are called ``aspect words''. Please refer to Appendix~\ref{abae_result} for examples.} For each review, the ``Percentage of Aspect Words'' represents the percentage of aspect words that appear in the review sentence. If the reviews contain more aspect words, it indicates that it has more hidden fine-grained aspect information, which may be more similar to what a human would write.
\end{itemize}

\noindent\textbf{Implementation Details.}
In the following, we provide the training details for each module in the proposed ARG.\footnote{https://github.com/hongyuntw/RBRS-ARG}
\begin{itemize}
    \item \textbf{ABAE}: When training ABAE, we choose $k$=15 for the number of hidden aspects in three datasets, and all other settings are the same as the original work.
    \item \textbf{Language Model}: We fine-tune distil-gpt2 released on huggingface\footnote{https://huggingface.co/distilgpt2}, set the batch size to 64, the maximum sequence length to 128, train 10 epochs, and select the one with the best validation loss as our evaluation model.
    \item \textbf{Attack Review Generator:} We choose bart-base-cnn\footnote{https://huggingface.co/ainize/bart-base-cnn} as our base generator. In the leave-one-out training stage, our dataset format is built according to \cite{Fewsum} and trained with a batch size of 64 for 3 epochs. For reinforcement learning, we train 5 epochs with a batch size of 64. The maximum generated review length is also set to 128 to fit our evaluation of LM, and the value of $\lambda$ is set to 0.5 based on experiments on the validation set. Finally, we choose the checkpoint that has the highest rewards in the validation set as our final generator. The relevance reward and the target aspect distribution selection are performed based on the ROUGE-1 score.
    \item \textbf{Fake user's ID embedding:} DeepCoNN and NRCMA do not consider the identity of the user who writes the item reviews. For the RMG, the fake user ID embedding is randomly initialized in our experiments.
\end{itemize}

\begin{table*}[!t]
    \renewcommand\arraystretch{1.2}
    \caption{Examples from the Musical Instruments and Office Products show the reviews generated under different settings of ARG and attack baselines against DeepCoNN. ``P'', ``I'', and ``R'' refer to reward prediction shift, inverse perplexity, and relevance in reinforcement learning, respectively. ``A'' refers to training with aspect generation. The blue sentences describe the detailed features of the product; the underlined words are the aspect words; the red words describe the attack keywords; and the orange words are the modification of TextBugger and HotFlip.}
    \footnotesize
    \begin{tabularx}{\textwidth}{m|s|g|b}
        \toprule
        Item & Attack Setting & PS / PPL & Generated Reviews \\

        \midrule                  
        \multirow{5}{*}[-1.0cm]{\shortstack{ Presentation Book \\ (Office Products)}} 
                            &  Copycat & 0.009 / 6.43 & Exactly what I was looking for,  I use it to hold my prints for shows. Just as nice as it looks, nothing more, nothing less.\\
        \cline{2-4}
                            & TextBugger & 0.009 / 6.43 & Exactly what I was looking for,  I use it to hold my prints for shows. Just as nice as it looks, nothing more, nothing less. \\
        \cline{2-4}
                            & HotFlip & 0.012 / 6.81 &   Exactly what I was looking for,  I use it to hold my prints for shows. Just as \sout{nice}\textcolor{orange}{perfect} as it looks, nothing more, nothing less.\\
        \cline{2-4}
        
                                &  P & 0.531 / 206.4 & Perfect Love Great Quality Great Beautiful Great Well Great Amazing Great Good Great Wonderful Great Highly Love Highly Recommend Highly Would Highly Buy Great This Great Top Great  \\
        \cline{2-4}
                                &  P + I & 0.398 / 4.98 & This is a \textcolor{red}{great} product. It is exactly what I was looking for. I am very happy with it and would recommend it to anyone. The price was \textcolor{red}{great} and it arrived in \textcolor{red}{perfect} condition. Thank you. \\
        \cline{2-4}
                                &  P + I + R & 0.240 / 6.92 & This is the best file folder I have seen in my life. I \textcolor{red}{love} it and I use it all the time. It is exactly what I was looking for and it is \textcolor{red}{perfect} for me. The quality of the file is very \textcolor{red}{good} and the fact that it a \textcolor{red}{great} thing to have. \\
        \cline{2-4}
                                &  P + I + R + A & 0.239 / 7.12&  This is a \textcolor{red}{great} file folder.  \textcolor{my_blue}{I use it to hold my \underline{prints} for shows.} \textcolor{my_blue}{It is \underline{sturdy}} and \textcolor{red}{perfect} for what I need. The price is \textcolor{red}{great}, and I \textcolor{red}{love} it. This is the \textcolor{red}{perfect} \textcolor{my_blue}{ \underline{size} for a good  \underline{amount} of photos}. You can 't go wrong with this one! \\
        \bottomrule        

    \end{tabularx}
    \label{tab:example_reviews}
    \vspace{-4mm}

\end{table*}

\subsection{RQ1: Attack Performance}
The columns of PS in Table~\ref{tab:main} compare the attack performance, i.e., prediction shift, of different methods with three RBRSs on datasets in three different domains of Amazon 5-star datasets, together with Yelp dataset. The results indicate that the prediction shift of DeepCoNN caused by ARG is at least 15 times greater than that of TextBugger. For NRCMA and RMG, the proposed ARG outperforms the baselines by at least 40.5\% and 82.6\% in terms of prediction shift on Amazon and Yelp datasets, respectively. In contrast, Copycat can only make a relatively small prediction shift. This is because highly rated reviews can express user preferences well and represent positive sentiment but do not always contain attack keywords. As such, it is difficult to shill output ratings for all candidate users.


TextBugger and HotFlip perform better than Copycat since they attempt to replace the words that have a greater impact on RBRS rather than using high-rating reviews. Nevertheless, the performance of TextBugger and HotFlip is still worse than that of ARG since they only replace words in existing reviews. Although HotFlip uses gradient search to find other words that can introduce a larger prediction shift, the word-level HotFlip is still constrained by the limited words in the reviews and needs to maintain the semantics of reviews after replacing words. Therefore, these modification-based text adversarial attacks can only have limited prediction shifts in this attack scenario.

Moreover, the attack performance varies depending on the RBRSs. As DeepCoNN uses only review embeddings during inference without any user or item embeddings, it is significantly affected when the attack reviews are injected in DeepCoNN. NRCMA and RMG are less affected than DeepCoNN because they predict ratings according to the ID embeddings of users and items in addition to review embeddings. Although user and item embeddings seem to reduce the impact of attack reviews on RBRS, it is difficult to integrate the latest reviews into these two embeddings unless retraining the model.

\subsection{RQ2: Quality of Attack Reviews}
\label{rq2}
In addition to the attack effectiveness, the attack reviews should also be of high quality. Therefore, we evaluate the quality of reviews in terms of fluency, product relevance, and aspect performance. Moreover, we provide two case studies to demonstrate the quality of generated attack reviews. 


\noindent\textbf{Fluency.} We first use the perplexity of the reviews in the testing set obtained by the LM to estimate the fluency (a smaller value indicates better fluency). Table~\ref{tab:main} shows that the perplexity of reviews generated by TextBugger is approximately 4-8 times higher than that generated by Copycat or HotFlip. This is because Copycat directly selects highly-rated reviews written by humans from the corpus to achieve the attack, leading to a similar perplexity with the in-corpus reviews. On the other hand, HotFlip replaces the words only if they have similar semantic meaning and have the same part-of-speech. These constraints make the attack review preserve the original meaning and ensure grammatical correctness. TextBugger performs the attacks by inserting/replacing/deleting the characters to candidate words or tokens in a review sentence. Therefore, there is a high probability that these modified words may become Out-of-Vocabulary (OOV) tokens, leading to a high perplexity by the LM. In contrast, benefiting from the inverse perplexity reward, the attack reviews generated by ARG obtained a low perplexity close to the reference value, indicating that those reviews are sufficiently fluent from the LM perspective. 
 
\noindent\textbf{Relevance.} The relevance scores in Table~\ref{tab:main} show that ARG significantly outperforms other baselines. The main reason is that the input of baselines is randomly selected among all the related 5-star reviews. As such, some 5-star reviews may only contain limited information. One alternative is to select the 5-star review with the highest relevance score as the input. However, this approach may lead to a low prediction shift and aspect generation since the common sentences among the reviews may be simple or short, \textit{e.g.}, ``This is good''. It is worth noting that Copycat and HotFlip are similar to the in-corpus reviews since Copycat selects the reviews from the corpus, while HotFlip replaces only one or two words in the reviews under the constraints. Due to the text modification, TextBugger may replace some words related to the product in the review sentences, resulting in a lower relevance score than that of Copycat. ARG benefits from the relevance reward in RL and can improve its score by capturing additional patterns beyond one review by integrating different information about the items from their historical reviews into the new attack reviews. Please note that the highly-relevant and fluent reviews generated by ARG may not be easily perceived as spam or meaningless information since these reviews provide a great deal of product information. They are more likely to be considered valuable reviews.

\begin{table}[t]
    \caption{Comparisons of human evaluation results.}
    \centering
    \begin{tabular}{c|c c}
     \toprule
     Reviews Type & Fluency & Informativeness \\
     \midrule
      Copycat  & 4.00 & 3.47\\
      \hline
      ARG w/o. aspect  & 3.94  & 3.56\\
      \hline
      ARG    &  \textbf{4.03} & \textbf{3.98}\\
      \bottomrule  
    \end{tabular}
    \label{tab:human}
    \vspace{-3mm}
\end{table}

\begin{table}[t]
        \caption{Percentage of reviews classified as human-written by a fake sentence classifier on the Amazon Musical Instruments domain."}
        \centering
        \small
        \begin{tabular}{cccc}
        \toprule 
        \multicolumn{2}{c}{ARG} & \multirow{2}{*}{FewSum} & \multirow{2}{*}{Human Written} \\
            RBRS & P / +I / +R / +A & &  \\
        \midrule
        DeepCoNN & 0\% / 0\% / 0\% / 16.8\% & \multirow{3}{*}{38.6\%} & \multirow{3}{*}{68.6\%} \\
        NRCMA &  12.4\% / 25.2\% / 39.2\% / 45\% &  &  \\
        RMG & 15.4\% / 19.8\% / 31.8\% / 42\%  &  &  \\
        \bottomrule
        \end{tabular}
        \label{classifier}
        \vspace{-3mm}
\end{table}



\noindent\textbf{Aspect Performance.} The aspect generation performance in Table~\ref{tab:main} shows that the proposed ARG outperforms the ARG without the aspect loss, indicating that aspect loss in the ARG is necessary for generating fine-grained information. Taking the product "Flat Bottom Strat Nut" in Table~\ref{tab:example_reviews} as an example, the review generated by ARG without aspect is \textit{``I am very happy with the quality of this nut''}, which simply mentions that the user is satisfied with the quality of the product. However, equipped with the aspect loss, ARG provides more details about the quality, such as \textit{``It is easy to set up, sturdy''}, and provides additional positive reasons with the product, such as \textit{``The price is great and this nut works as well as any other nut I have ever used.''}. Moreover, the aspect performance of ARG is also higher than that of in-corpus reviews since the reviews in the corpus are written by a single person with limited information. This is because the reviews generated by ARG are composed of specific product information from multiple reviews, which allows ARG to easily generate more aspect words.

\noindent\textbf{Case Study.} Table~\ref{tab:example_reviews} shows the different attack reviews of baselines and ARGs. In the example, TextBugger is unable to find a suitable word to create a higher prediction shift compared to Copycat, while HotFlip still successfully replaces the word \textit{nice} with \textit{perfect} to produce a greater prediction shift. The example demonstrates that the modification-based approach may be limited by the input review. In contrast, ARGs generate reviews containing multiple attack keywords while maintaining review fluency. Due to the space constraint, please refer to Appendix~\ref{more_case_study} for more examples.

\begin{table*}[t]
\caption{The performance of recommendation and shilling attack before and after adversarial training (AT) on Amazon Musicial Instruments dataset. ``C'' refers to Copycat, ``T'' refers to TextBugger, and ``H'' refers to HotFlip.}
    \centering
    \footnotesize
    \begin{tabular}{c|c|c|ccc|cccc}
    \toprule 
    \multirow{3}{*}{RBRS} & \multirow{3}{*}{\shortstack{Adversarial \\ Samples}} & \multirow{3}{*}{MSE} &\multicolumn{7}{c}{Prediction shift of attacks} \\ 
    \cline{4-10}
        & & & \multicolumn{3}{c|}{Attack Baselines} & \multicolumn{4}{c}{ARG} \\
        & & & C & T & H & P & P+I & P+I+R & P+I+R+A  \\
    \midrule
    \multirow{5}{*}{DeepCoNN} & None & 0.868 & 0.012 & 0.017 & 0.035 & 0.670 & 0.619 & 0.391 & 0.363 \\
    & \makecell[c]{P} & 0.867 & \shortstack{0.0004  (-96.66\%)} & \shortstack{0.0002  (-98.41\%)} & \shortstack{ 0.0004 (-98.85\%)} & \shortstack{0.0000  (-99.99\%)} & \shortstack{0.0035  (-99.42\%)} & \shortstack{0.0014  (-99.62\%)} & \shortstack{0.0003 (-99.90\%)}\\
    & \makecell[c]{P + I} & 0.867 & \shortstack{0.0011  (-90.66\%)} & \shortstack{0.0030  (-82.35\%)} & \shortstack{0.004 (-88.57\%)} & \shortstack{0.326 (-51.34\%)} & \shortstack{0.004 (-99.35\%)} & \shortstack{0.008 (-97.95\%)} & \shortstack{0.004 (-98.89\%)}\\
    & \makecell[c]{P + I + R} & 0.868 & \shortstack{0.0028 (-75.84\%)} & \shortstack{0.0030 (-82.25\%)} & \shortstack{0.002 (-94.28\%)} & \shortstack{0.281 (-58.01\%)} & \shortstack{0.047 (-92.35\%)} & \shortstack{0.019 (-95.02\%)} & \shortstack{0.064 (-82.10\%)}\\
    & \makecell[c]{P + I + R + A} & 0.866 & \shortstack{0.001 (-91.23\%)} & \shortstack{0.0009 (-94.47\%)} & \shortstack{0.007 (-80.00\%)} & \shortstack{0.179 (-73.16\%)} & \shortstack{0.0128 (-97.92\%)} & \shortstack{0.028 (-92.58\%)} & \shortstack{0.005 (-98.38\%)}\\
    \hline
    \multirow{5}{*}{NRCMA} & None & 0.861 & 0.027 & 0.043 & 0.055 & 0.157 & 0.048 & 0.077 & 0.084 \\
    & \makecell[c]{P} & 0.860 & \shortstack{0.0019  (-92.96\%)} & \shortstack{0.0002  (-99.53\%)} & \shortstack{ 0.0014 (-99.86\%)} & \shortstack{0.0039  (-97.51\%)} & \shortstack{0.0002  (-99.50\%)} & \shortstack{0.0026  (-96.62\%)} & \shortstack{0.0027 (-96.78\%)}\\
    & \makecell[c]{P + I} & 0.858 & \shortstack{0.0021  (-91.86\%)} & \shortstack{0.0003  (-99.20\%)} & \shortstack{0.0017 (-96.82\%)} & \shortstack{0.0023 (-98.53\%)} & \shortstack{0.0001 (-99.87\%)} & \shortstack{0.00004 (-99.93\%)} & \shortstack{0.00007 (-99.91\%)}\\
    & \makecell[c]{P + I + R} & 0.862 & \shortstack{0.0019 (-92.62\%)} & \shortstack{0.0007 (-98.22\%)} & \shortstack{0.0011 (-97.84\%)} & \shortstack{0.0001 (-99.90\%)} & \shortstack{0.00003 (-99.92\%)} & \shortstack{0.0002 (-99.71\%)} & \shortstack{0.0001 (-99.78\%)}\\
    & \makecell[c]{P + I + R + A} & 0.861 & \shortstack{0.002 (-92.59\%)} & \shortstack{0.0005 (-98.83\%)} & \shortstack{0.0019 (-96.54\%)} & \shortstack{0.000 (-99.99\%)} & \shortstack{0.0000 (-99.99\%)} & \shortstack{0.0015 (-97.93\%)} & \shortstack{0.0012 (-98.49\%)}\\
   \hline
    \multirow{5}{*}{RMG} & None & 0.860 & 0.014 & 0.031 & 0.055 & 0.155 & 0.092 & 0.089 & 0.091 \\
    & \makecell[c]{P} & 0.861 & \shortstack{0.0052  (-62.73\%)} & \shortstack{0.0109  (-64.83\%)} & \shortstack{ 0.0123 (-77.55\%)} & \shortstack{0.0539  (-65.29\%)} & \shortstack{0.0356  (-61.32\%)} & \shortstack{0.0343  (-61.37\%)} & \shortstack{0.0283 (-68.85\%)}\\
    & \makecell[c]{P + I} & 0.859 & \shortstack{0.0045  (-67.22\%)} & \shortstack{0.0113  (-63.49\%)} & \shortstack{0.0071 (-86.94\%)} & \shortstack{0.0610 (-60.71\%)} & \shortstack{0.0256 (-72.21\%)} & \shortstack{0.0310 (-65.90\%)} & \shortstack{0.0352 (-61.28\%)}\\
    & \makecell[c]{P + I + R} & 0.859 & \shortstack{0.0040 (-71.22\%)} & \shortstack{0.0098 (-68.38\%)} & \shortstack{0.0056 (-89.77\%)} & \shortstack{0.0588 (-62.16\%)} & \shortstack{0.0256 (-72.17\%)} & \shortstack{0.0307 (-65.43\%)} & \shortstack{0.0316 (-65.22\%)}\\
    & \makecell[c]{P + I + R + A} & 0.858 & \shortstack{0.0051 (-92.90\%)} & \shortstack{0.0117 (-62.20\%)} & \shortstack{0.0057 (-89.62\%)} & \shortstack{0.0600 (-61.37\%)} & \shortstack{0.0266 (-71.02\%)} & \shortstack{0.0290 (-67.41\%)} & \shortstack{0.0298 (-67.24\%)}\\

    \bottomrule
    \end{tabular}
    \label{tab:adversarial}
    \vspace{-3mm}
\end{table*}

\subsection{RQ3: Human Evaluation and Detector}
For human evaluation, six products from each domain are selected for promotion. For each product, we generate three types of reviews in different ways, including Copycat, ARG without aspect generation, and ARG, and ask people to rate these reviews. We recruited 53 users from the university, where 12 participated users had research experience in NLP/deep learning and the rest of the users were not familiar with NLP/deep. The users are asked to answer the following two Likert scale survey questions: 1) fluency: the review should be easy to read and grammatically correct, and; 2) informativeness: the review should contain more information about the product. The ratings range from 1 to 5, and a higher score indicates a better result.

Table~\ref{tab:human} shows that the fluency scores are similar for the reviews generated by different methods. The results indicate that the reviews generated by ARG are inclined to be grammatically correct and easy to read as reviews written by humans. For informativeness, the results are consistent with RQ2. Additionally, the ARG surpasses the former two in informativeness scores since it can add more granular information through aspect generation.

In addition to human evaluation, we introduce an open-source fake sentence classifier, roberta-base-openai-detector\footnote{https://huggingface.co/roberta-base-openai-detector}, which can be used to detect text generated by GPT-2 models. Table~\ref{classifier} shows the percentage of different types of reviews that are classified as human-written by the classifier. We sampled 500 reviews as the input and compared the reviews generated by ARG with human reviews and reviews generated by FewSum~\cite{Fewsum}, which is a state-of-the-art review summarizer. The results show that 1) the relevance rewards can make the reviews more human-like on NRCMA and RMG, while aspect generation is advantageous for making the reviews more human-like on three different RBRSs. 2) Although the results show that the classifier considers the reviews generated by ARG to be not quite like those written by humans, the classifier does not perform well on actual human-written reviews (only 68.6\%). This indicates that detecting AI-generated sentences is challenging, while the reviews generated by ARG may not be easily detected. We also perform similar classification experiments with ChatGPT~\footnote{https://openai.com/blog/chatgpt/}. Due to the space constraint, please refer to Appendix~\ref{chatGPT} for more details.

\vspace{-1.5mm}
\subsection{RQ4: Adversarial Training}

To understand the performance of the RBRS before and after adversarial training, we add the generated adversarial reviews under various reward settings into the training dataset and train on all three RBRSs~\footnote{ARGs are used to generate attack reviews for each item within the training set. Subsequently, these attack reviews are included in the review set of the corresponding item. The RBRSs are then trained on the dataset including attacked reviews with the objective of predicting the ground truth in the presence of malicious attacks.}. Table~\ref{tab:adversarial} shows the recommendation performance, together with the shilling attack performance before and after adversarial training on the Amazon Musical Instruments domain. Due to space limitations, the results of adversarial training on Yelp can be found in Appendix~\ref{yelp-adv}. Equipped with the adversarial samples generated by ARG, the results show that the retrained RBRSs are able to effectively defend against Copycat and TextBugger. Moreover, the adversarial training also helps DeepCoNN, NRCMA and RMG to defend against ARGs with various reward configurations by at least 51.34\%, 91.86\% and 60.71\%, respectively. This is because adversarial reviews capture spurious keywords that can easily influence the rating score. Therefore, training on the reviews generated by ARG encourages the RBRS to learn more robust features instead of relying on a simple correlation. In addition, the model still maintains the MSE performance after retraining. In summary, the proposed ARG can act as the weakness scanner of the RBRS, and training on these adversarial examples can make the RBRS more robust to different kinds of attacks.

\subsection{Ablation Study}
\begin{table}[t]
\caption{Attack performance of ARGs trained with different rewards against DeepCoNN on the Musical Instruments domain.}
    \centering
    \begin{tabular}{c|ccc}
        \toprule
        Rewards & PS & PPL & Relevance \\
        \midrule
            $PS$ & \textbf{0.670} & 150.31 & 0.081 \\
            $PS + InvPPL$ & 0.619 & \textbf{6.87} & 0.198 \\
            $PS + InvPPL + Relevance$ & 0.391 & 8.32 & \textbf{0.253} \\
        \bottomrule  
    \end{tabular}
    \label{tab:ablation-study-rewards}
    \vspace{-5mm}

\end{table}

To examine the effects of each reward, we first conduct the ablation study in this section. Table~\ref{tab:ablation-study-rewards} shows that the attack review generated by ARG with only the ${PS}$ reward can generate the largest prediction shift to shill the RBRS. However, the text quality of the generated reviews is the worst. Table~\ref{tab:example_reviews} also shows an example of generated reviews under different training settings. The reviews contain only repetitive attack keywords if we only take ${PS}$ as the reward, which can be easily detected and has the largest perplexity and lowest relevance score. Moreover, when shilling different domains of RBRS, the attack keywords will be different. In the Musical Instruments domain, the keywords for attack are \textit{perfect}, \textit{love} and \textit{awesome}. In the Office Products domain, the attack words are more diverse, such as \textit{Wonderful} and \textit{Highly}, but all of them represent positive sentiment and are effective in promoting products.

Moreover, if both ${InvPPL}$ and ${PS}$ are used as rewards, ${InvPPL}$ imposes a constraint on the ARG to make the generated reviews more fluent. However, the generated reviews under this setting cannot capture the information about the product, leading to similar reviews for all products. Table~\ref{tab:example_reviews} shows that the reviews generated under this setting are simply praising the products by having offensive keywords. Finally, by introducing $Relevance$ to the rewards, the generated reviews can capture the essential information about the product. As shown in Table~\ref{tab:ablation-study-rewards}, the relevance scores are higher than in the previous two settings. Although the prediction shift is slightly lower than theirs, the generated reviews strike a better balance between attack effectiveness and quality. Table~\ref{tab:example_reviews} also shows that the reviews can acquire information such as the product name (nut and file folder) and common attributes in its historical reviews.

\section{Ethical Considerations}
The method presented in this study has the potential to be misused to infiltrate review-based recommender systems with fake reviews that are challenging to identify and eliminate. However, the goal is not to cause harm, but rather to expose these vulnerabilities to the public in a similar manner to how ethical hackers reveal security flaws. Additionally, we also proposed an adversarial training method to defend against such attacks. Consequently, this research ultimately serves a long-term benefit for the community. Throughout the course of this study, we adhered to the ethical guidelines established by the ACM to minimize any harm and all the experiments were conducted on simulated recommender systems rather than real-world recommender systems.

\section{Conclusion}
In this paper, we discover a new vulnerability in RBRS caused by a shilling attack. To generate a malicious review that can effectively deceive RBRSs with high fidelity, we propose the Attack Review Generator (ARG), which leverages a reinforcement learning framework with three auxiliary rewards and aspect generation training to encourage the ARG to generate fine-grained information about the product. Experimental results demonstrate that the reviews generated by the ARG are effective for attacking different RBRSs while maintaining high quality. Human evaluation further shows that our reviews can provide more detailed information than regular reviews. Most importantly, equipped with the reviews generated by the ARG and adversarial training, the RBRS can be much more robust to malicious reviews generated by different attack methods. In the future, we plan to study shilling attacks with multiple reviews.

\section*{Acknowledgement}
\label{sec:ack} 
This work was supported in part by the National Science and Technology Council of Taiwan under Grants NSTC-109-2221-E-009-114-MY3 and NSTC-111-2221-E-001-021.

\newpage


\bibliographystyle{ACM-Reference-Format}
\balance
\bibliography{main}


\appendix
\section{Appendix}

\begin{figure}[h]
    \centering
    \begin{subfigure}{\linewidth}
        \includegraphics[width=\linewidth]{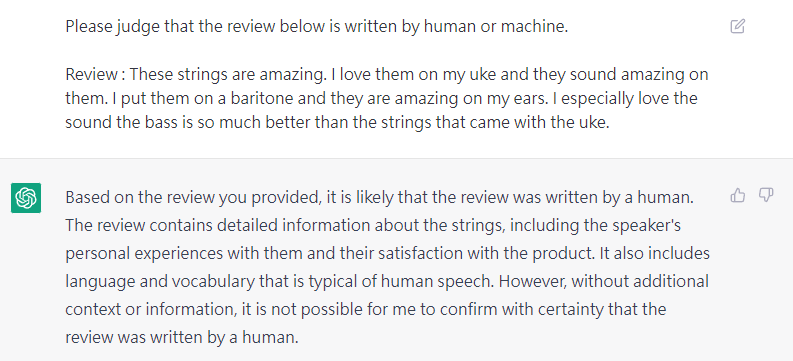}
        \caption{An example of ChatGPT's classification of a review generated by ARG as human-written or machine-generated.}
        \label{fig:chatgpt-single}
    \end{subfigure}
    \begin{subfigure}{\linewidth}
        \includegraphics[width=\linewidth]{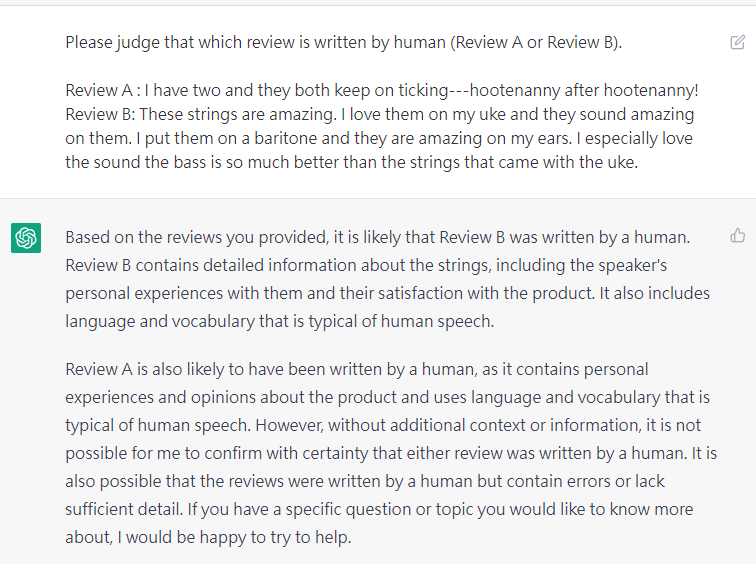}
        \caption{An example of ChatGPT's classification between human written reviews and reviews generated by ARG.}
        \label{fig:chatgpt-second}
    \end{subfigure}
    \caption{Examples of AI-written text detection by ChatGPT.}
    \label{chatGPT-results-fig}
    \vspace{-5mm}
\end{figure}

\begin{table}[h]
    \caption{List of representative words for inferred aspects on three domains.}
    \scriptsize
    \begin{tabular}{c | c}
        \toprule
        Inferred Aspects & Representative Words \\
        \midrule
        & \textbf{Musical Instrument} \\
        \hline
        Performance & performs, functioned, executed, constructed, serves, works \\
        Use & use, setup, operate, usage, access, hookup, \\
        Quality & quality, value, sturdy, solidly, fidelity, durable \\
        Electronic connections & cable, cord, shorting, adaptors, connectors \\
        Appearance & colors, translucent, looks, aluminum, metallic, illumination \\
        \hline
        & \textbf{Video Games} \\
        \hline
        Computer & firewall, windows, uninstalling, reinstalling, startup \\
        Unboxing & picked, opened, got, excited, arrived, received \\
        Motion & walked, shoot, hovering, chasing, flying, running \\
        Family &  son, daughter, husband, grandson, brother, fiance, nephew \\
        Time &  years, months, decade, week, days, semesters \\
        \hline
        & \textbf{Office Products} \\
        \hline
        Printer & printer, canon, epson, print, cartridges, lexmark, toner, ink \\
        Paper & paper, sheets, cardstocks, vellum, linen, envelopes, newsprint \\
        Texture & solid, flimsy, sturdy, robust, spongy \\
        Amounts & several, many, numerous, dozens, size, amount, various \\
        \bottomrule
    \end{tabular}
    \label{tab:abae}
\end{table}

\begin{table*}[h]
\caption{The performance of recommendation and shilling attack before and after adversarial training (AT) on Yelp dataset. ``C'' refers to Copycat, ``T'' refers to TextBugger, and ``H'' refers to HotFlip.}
    \centering
    \footnotesize
    \begin{tabular}{c|c|c|ccc|cccc}
    \toprule 
    \multirow{3}{*}{RBRS} & \multirow{3}{*}{\shortstack{Adversarial \\ Samples}} & \multirow{3}{*}{MSE} &\multicolumn{7}{c}{Prediction Shift of Attacks} \\ 
    \cline{4-10}
        & & & \multicolumn{3}{c|}{Attack Baselines} & \multicolumn{4}{c}{ARG} \\
        & & & C & T & H & P & P+I & P+I+R & P+I+R+A  \\
    \midrule
    \multirow{5}{*}{DeepCoNN} & None & 1.171 & 0.029 & 0.023 & 0.030 & 1.222 & 0.207 & 0.233 & 0.228 \\
    & \makecell[c]{P} & 1.170 & \shortstack{0.0035  (-87.93\%)} & \shortstack{0.0078  (-66.09\%)} & \shortstack{ 0.0039 (-87.00\%)} & \shortstack{0.0062  (-99.49\%)} & \shortstack{0.0288  (-86.09\%)} & \shortstack{0.0297  (-87.25\%)} & \shortstack{0.0286 (-87.46\%)}\\
    & \makecell[c]{P + I} & 1.172 & \shortstack{0.0036  (-87.59\%)} & \shortstack{0.0022  (-90.43\%)} & \shortstack{0.0038 (-87.33\%)} & \shortstack{0.0198 (-98.38\%)} & \shortstack{0.0025 (-98.79\%)} & \shortstack{0.0248 (-89.36\%)} & \shortstack{0.0233 (-89.78\%)}\\
    & \makecell[c]{P + I + R} & 1.171 & \shortstack{0.0021 (-92.76\%)} & \shortstack{0.0032 (-86.09\%)} & \shortstack{0.0021 (-93.00\%)} & \shortstack{0.1961 (-83.96\%)} & \shortstack{0.0029 (-98.60\%)} & \shortstack{0.0078 (-96.65\%)} & \shortstack{0.0061 (-97.32\%)}\\
    & \makecell[c]{P + I + R + A} & 1.169 & \shortstack{0.0030 (-89.66\%)} & \shortstack{0.0070 (-69.57\%)} & \shortstack{0.0031 (-89.67\%)} & \shortstack{0.0204 (-98.33\%)} & \shortstack{0.0260 (-87.44\%)} & \shortstack{0.0197 (-91.55\%)} & \shortstack{0.0107 (-95.31\%)}\\
    \hline
    \multirow{5}{*}{NRCMA} & None & 1.129 & 0.033 & 0.041 & 0.052 & 0.319 & 0.102 & 0.095 & 0.087 \\
    & \makecell[c]{P} & 1.128 & \shortstack{0.0032  (-90.30\%)} & \shortstack{0.0025  (-93.90\%)} & \shortstack{ 0.0028 (-94.62\%)} & \shortstack{0.0028  (-99.12\%)} & \shortstack{0.0035  (-96.57\%)} & \shortstack{0.0040  (-95.79\%)} & \shortstack{0.0032 (-96.32\%)}\\
    & \makecell[c]{P + I} & 1.127 & \shortstack{0.0040  (-87.88\%)} & \shortstack{0.0023  (-94.39\%)} & \shortstack{0.0039 (-92.50\%)} & \shortstack{0.0036 (-98.87\%)} & \shortstack{0.0033 (-96.76\%)} & \shortstack{0.0034 (-96.42\%)} & \shortstack{0.0034 (-96.09\%)}\\
    & \makecell[c]{P + I + R} & 1.130 & \shortstack{0.0041 (-87.58\%)} & \shortstack{0.0028 (-93.17\%)} & \shortstack{0.0052 (-90.00\%)} & \shortstack{0.0023 (-99.28\%)} & \shortstack{0.0043 (-95.78\%)} & \shortstack{0.0024 (-97.47\%)} & \shortstack{0.0042 (-95.17\%)}\\
    & \makecell[c]{P + I + R + A} & 1.128 & \shortstack{0.0031 (-90.61\%)} & \shortstack{0.0037 (-90.98\%)} & \shortstack{0.0060 (-88.46\%)} & \shortstack{0.0041 (-98.71\%)} & \shortstack{0.0042 (-95.88\%)} & \shortstack{0.0025 (-97.37\%)} & \shortstack{0.0029 (-96.67\%)}\\
   \hline
    \multirow{5}{*}{RMG} & None & 1.133 & 0.0226 & 0.0239 & 0.0261 & 0.1187 & 0.0464 & 0.0635 & 0.0797 \\
    & \makecell[c]{P} & 1.132 & \shortstack{0.0055  (-75.66\%)} & \shortstack{0.0063  (-73.64\%)} & \shortstack{ 0.0060 (-77.01\%)} & \shortstack{0.0027  (-97.73\%)} & \shortstack{0.0049  (-89.44\%)} & \shortstack{0.0033  (-94.80\%)} & \shortstack{0.0048 (-93.98\%)}\\
    & \makecell[c]{P + I} & 1.129 & \shortstack{0.0064  (-71.68\%)} & \shortstack{0.0068  (-71.55\%)} & \shortstack{0.0075 (-71.26\%)} & \shortstack{0.0028 (-97.64\%)} & \shortstack{0.0048 (-89.66\%)} & \shortstack{0.0043 (-93.23\%)} & \shortstack{0.0043 (-94.60\%)}\\
    & \makecell[c]{P + I + R} & 1.131 & \shortstack{0.0045 (-80.09\%)} & \shortstack{0.0049 (-79.50\%)} & \shortstack{0.0057 (-78.16\%)} & \shortstack{0.0028 (-97.64\%)} & \shortstack{0.0048 (-89.66\%)} & \shortstack{0.0063 (-90.08\%)} & \shortstack{0.0047 (-94.10\%)}\\
    & \makecell[c]{P + I + R + A} & 1.132 & \shortstack{0.0036 (-86.28\%)} & \shortstack{0.0027 (-88.70\%)} & \shortstack{0.0069 (-73.56\%)} & \shortstack{0.0032 (-97.30\%)} & \shortstack{0.0064 (-86.21\%)} & \shortstack{0.0060 (-90.55\%)} & \shortstack{0.0068 (-91.47\%)}\\

    \bottomrule
    \end{tabular}
    \label{tab:adversarial-yelp}
\end{table*}

\subsection{Results of ABAE}
\label{abae_result}
Table~\ref{tab:abae} shows five inferred aspects extracted by ABAE and representative words in each domain. Please note that the inferred aspects are manually labeled.

\subsection{Adversarial Training Results on Yelp Dataset}
\label{yelp-adv}

Table~\ref{tab:adversarial-yelp} illustrates the change in prediction shifts before and after incorporating adversarial training, which is similar to those found in the Amazon dataset. Using adversarial training with reviews generated by ARG, DeepCoNN, NRCMA, and RMG, can significantly decrease the impact of all types of attacks, by at least 66.09\%, 87.58\%, and 71.26\%, respectively, while still maintaining the same level of recommendation performance.

\subsection{ChatGPT as Fake Review Classifier}
\label{chatGPT}

We leverage ChatGPT~\footnote{https://openai.com/blog/chatgpt/} as a tool for detecting AI-generated reviews. We conducted two kinds of experiments. The first was to ask ChatGPT to determine if a review was written by a human or machine. In the second kind of experiment, we presented ChatGPT with two types of reviews, one written by a human and the other one generated by ARG, and asked ChatGPT to determine which one was more likely to be written by a human. For both experiments, we randomly select 50 reviews generated by ARG as the query inputs. Fig~\ref{chatGPT-results-fig} shows examples of these two kinds of experiments. For the first experiment, the results show that ChatGPT answered that 49 out of 50 reviews were written by a human, and provided the reason that the reviews contain detailed information about the products. For the second experiment, the results showed that ChatGPT considered both the reviews generated by ARG and human-written reviews as human-written in 46 out of 50 cases. This finding indicates that the reviews generated by ARG are highly similar to human-written reviews, making it difficult to distinguish between the two. It is worth noting that since ChatGPT is not specifically designed for detecting machine-generated reviews, its capability in this area may be limited. Therefore, we only put this experiment as an additional analysis in the appendix.


\subsection{Examples of Different Attack Reviews}
\label{more_case_study}

Table~\ref{tab:example_reviews_appendix} shows another example of attack reviews generated by different baselines and ARGs. In the example, TextBugger can generate an attack review with a minor prediction shift from Copycat by creating a ``bug'' in the word \textit{again} by replacing it with the Out-of-Vocabulary term \textit{aga in}. However, this modification results in a high perplexity for the review. In contrast, HotFlip replaces the word \textit{great} with \textit{excellent} to produce a larger prediction shift while preserving the semantic meaning. On the other hand, ARGs generate reviews that contain multiple attack keywords and maintain the fluency of the reviews.

\begin{table*}[!h]
    \renewcommand\arraystretch{1.2}
    \caption{Examples from the Musical Instruments and Office Products show the reviews generated under different settings of ARG and attack baselines against DeepCoNN. ``P'', ``I'', and ``R'' refer to reward prediction shift, inverse perplexity, and relevance in reinforcement learning, respectively. ``A'' refers to training with aspect generation. The blue sentences describe the detailed features of the product; the underlined words are the aspect words; the red words describe the attack keywords; and the orange words are the modification of TextBugger and HotFlip.}
    \footnotesize
    \begin{tabularx}{\textwidth}{m|s|g|b}
        \toprule
        Item & Attack Setting & PS / PPL & Generated Reviews \\

        \midrule
        \multirow{5}{*}[-1.3cm]{\shortstack{ Flat Bottom Strat Nut \\ (Musical Instrument)}} 
                                
                                &  Copycat & 0.015 / 7.01 &  Very nice. I installed on my guitar. My guitar stays in tune. Like that it is lubricated. Will buy again. Fits great and sounds great! every guitar should come with this nut \\
        \cline{2-4}
                                & TextBugger & 0.016 / 36.63 & Very nice. I installed on my guitar. My guitar stays in tune. Like it is lubricated. Will buy \textcolor{orange}{aga in}. Fits great and sounds great! every guitar should come with this nut \\
        \cline{2-4}
                                & HotFlip & 0.017 / 8.01  & Very nice. I installed on my guitar. My guitar stays in tune. Like that it is lubricated. Will buy again. Fits \sout{great}\textcolor{orange}{excellent} and sounds great! every guitar should come with this nut \\

        \cline{2-4}             
                                &  P & 0.674 / 112.5 & Perfect perfect perfect perfectly perfectly perfect love love perfect awesome love awesome awesome perfect gorgeous love perfectly awesome great awesome amazing awesome fantastic awesome! \\

        \cline{2-4}
                                &  P + I & 0.603 / 9.51  & This is a \textcolor{red}{perfect} product. It is \textcolor{red}{perfect} for my needs. I \textcolor{red}{love} it! I am very happy with this purchase. Thank you for offering this product \\
        \cline{2-4}
                                &  P + I + R & 0.383 / 8.41  & This is a great nut for the guitar. It fits \textcolor{red}{perfectly}. I \textcolor{red}{love} it. The sound is \textcolor{red}{perfect} for my guitar and I am very happy with the quality of this nut. \\
        \cline{2-4}
                                &  P + I + R + A & 0.386 / 8.12  & This nut is \textcolor{red}{perfect} for my guitar. I \textcolor{red}{love} the sound of it and it works great. It is \textcolor{my_blue}{easy to \underline{setup}, \underline{sturdy}} and I like it so much that I want to buy another one. The price is \textcolor{red}{great} and this nut \textcolor{my_blue}{\underline{works} as well as any other nut I have ever used.} \\

        \bottomrule            
    
    \end{tabularx}
    \label{tab:example_reviews_appendix}
    \vspace{-4mm}

\end{table*}

\end{CJK*}
\end{document}